\documentstyle[11pt,paspconf,psfig]{article}
\begin{document}

\title{Lyman $\alpha$ Absorption as a Sensitive Probe of the H~I
  Column in Cooling Flows}

\author{A. Laor}
\affil{Physics Department, Technion, Haifa 32000, Israel}

\begin{abstract}
X-ray spectra of a significant fraction of cooling flow (CF) clusters of 
galaxies indicate the presence of large columns of ``cold'' absorbing
gas. The physical nature of the absorbing medium remains a mystery. 
Searches for H~I absorption using the 21~cm hyperfine structure
line yielded null results in most cases. The purpose of this contribution 
is to
point out that the Lyman $\alpha$ absorption cross section is $\ge 10^7$ 
times larger than for the 21~cm line, it can therefore be used as a 
very sensitive probe of the H~I column in clusters, and can thus
place stringent constraints on the nature of the X-ray absorber. 
This method is applied to the Perseus CF cluster where a medium resolution
($\sim 250$~km~s$^{-1}$) UV spectrum is available. The upper limit on the 
H~I column obtained using Lyman $\alpha$ is at least
$\sim 50$ times smaller than the 21~cm detection, and $\sim 5,000$ smaller
than implied by X-ray spectra, indicating that the X-ray absorber is exceedingly
devoid of H~I. Higher resolution
UV spectra with HST may improve the H~I column limits by an additional 
factor of $\sim 4,000$. 
This method can be applied to strongly constrain the nature of the
X-ray absorbing medium in a significant fraction of CF clusters .
\end{abstract}


\section{Introduction}
The presence of ubiquitous X-ray absorption in cooling flow (CF) clusters
was first discovered by White et al. (1991) using the {\em Einstein}
Solid State Spectrometer (SSS), and was later verified using 
{\em EXOSAT},
{\em ROSAT}, and {\em ASCA} (e.g. Allen et al. 1992; Fabian et al.
1994). The typical observed columns are a few $10^{20}$ to a few 
$10^{21}$~cm$^{-2}$. Searches were conducted at other wavelengths in order
to verify the presence of the absorber and to understand its 
nature. Emission line constraints rule out absorbing gas at
$T\sim 10^5-10^6$~K, and thus the absorbing gas must be mostly
H~I and/or H$_2$. Extensive searches for H~I absorption using the
21~cm hyperfine structure line (e.g. Jaffe 1990; Dwarakanath, van Gorkom, 
\& Owen 1994) and searches for
CO associated with H$_2$ (e.g. O'dea et al. 1994; Antonucci \& Barvainis
1994) yielded upper limits typically well below
the X-ray columns. The 21~cm limits are linear with the
electron excitation temperature, and may thus be subject to significant
uncertainty.

For example, in the case of NGC~1275, the central galaxy in the
Perseus CF cluster, the X-ray column obtained by the {\em Einstein} SSS
is $1.3^{+0.3}_{-0.3}\times 10^{21}$~cm$^{-2}$ (White et al. 1991), by
{\em EXOSAT} $1.5^{+2.1}_{-0.9}\times 10^{21}$~cm$^{-2}$ 
(Allen et al. 1991), and by {\em ASCA} 
$3-4\times 10^{21}$~cm$^{-2}$ (Fabian et al. 1994). 
The 21~cm line, however, indicates an H~I 
column of only $2\times 10^{18}T_s$~cm$^{-2}$ (Jaffe 1990).  

In this paper we show that significantly tighter limits on 
$N_{\rm H~I}$ can be obtained using Ly$\alpha$. 
We first make a short comparison
of the absorption properties of Ly$\alpha$ versus the 21~cm line. We then
apply the results of curve of growth analysis for Ly$\alpha$ to 
show that the H~I column $\sim 10-20$~kpc away from the center 
of NGC~1275 is significantly
smaller than indicated by the 21~cm line, and demonstrate that improved
limits can be obtained with a higher quality UV spectrum.
We end with a short discussion of the implications of
the new upper limits on $N_{\rm H~I}$ on 
the nature of the X-ray absorber.

\section{On Lyman $\alpha$ vs. 21 cm Absorption}

In this section we compare the absorption properties of Lyman $\alpha$
versus the 21~cm line. In a two level atom the absorption cross section 
per atom is:
\[ \sigma_{\nu}=\frac{\pi e^2}{m_ec}f_{12}\phi(\nu)
f_{se}\frac{n_1}{n_1+n_2} \]
where $f_{12}$ is the oscillator strength, and
$\phi(\nu)$ is the line profile function (Voigt function for pure thermal
broadening). The parameter  
$f_{se}$ is the correction for stimulated emission given by
\[ f_{se}\equiv1-\frac{n_2}{n_1}\frac{g_1}{g_2} \]
where $n_i, g_i$ are the population and degeneracy of level i.
The level population is accurately described by the Boltzmann ratio
since collisions dominate both excitations and deexcitaions, and thus
\[ f_{se}=1-e^{-\frac{\Delta E}{kT}}, \]
or $f_{se}\simeq \frac{\Delta E}{kT}$ for $\Delta E\ll kT$.
The value of the line profile function at line center is  
\[ \phi(\nu_0)=\frac{1}{\sqrt{\pi}\Delta\nu_D}\] 
assuming a Gaussian line shape (i.e. thermal broadening), 
where $\Delta\nu_D=\nu_0\frac{b}{c}$ is the line width and 
$b=\sqrt{\frac{2kT}{m_p}}$ is the Doppler parameter 
(Rybicki \& Lightman 1979).

Table 1 compares the values of the parameters discussed above for
Ly$\alpha$ versus the 21~cm line.
\begin{table}
\caption{Ly$\alpha$ versus 21~cm absorption line parameters} 
\begin{center}
\begin{tabular}{ccc}
Parameter & Ly$\alpha$ & 21~cm\\
\tableline
$f_{12}$ & 0.416 & $5.75\times 10^{-12}$ \\
$\phi(\nu_0)$ & $5.33\times 10^{-10}T^{-1/2}$ & $9.27\times 10^{-4}T^{-1/2}$\\
$f_{se}$ & $\simeq 1$ & $0.0682T^{-1}$\\
$n_2/n_1$ & $\ll 1$ & 3 \\
$ \sigma_{\nu_0}$ & $5.88\times10^{-12} T^{-1/2}$ 
& $2.41\times10^{-19} T^{-3/2}$ \\
\end{tabular}
\end{center}
\end{table}
The ratio of line center absorption cross sections is therefore
\[  \frac
{\sigma_{\nu_0}({\rm Ly}\alpha)}
{\sigma_{\nu_0}(21 {\rm cm})}
=2.44\times 10^6 T,
\]
and since $T\ge 2.73$~K, Ly$\alpha$ is $10^7$ times more
sensitive to H~I absorption than the 21~cm line, if both absorption
lines are optically thin. 
Note that when $N_{\rm H~I}>2\times 10^{11}T^{1/2}$,
the Ly$\alpha$ line becomes optically thick, and the absorption 
equivalent width $EW=\int (1-e^{-\tau_{\nu}})d\nu$ increases only as
$\sqrt{\ln \tau}$.

Figure 1 presents a comparison of the absorption profiles of the 21~cm
line vs. Ly$\alpha$ for $10^{20}\ge N_{\rm H~I}\ge 10^{17}$~cm$^{-2}$.
Note the large difference in absorption EW of the two lines. 
\begin{figure}
\psfig{file=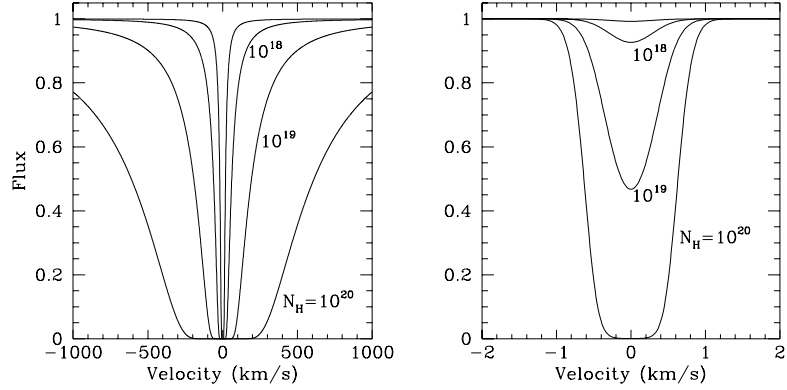,width=11.cm,angle=-90,silent=}
\caption{A comparison of the absorption profiles of the 21~cm
line (right) vs. Ly$\alpha$ (left) 
for $10^{17}\ge N_{\rm H~I}\ge 10^{20}$~cm$^{-2}$.
Note the large difference in velocity scales in the two panels.} 
\end{figure}
Jaffe (1990) measured in NGC~1275 21~cm absorption with 
$N_{\rm H~I}=2\times 10^{18}T$ 
and FWHM=477~km~s$^{-1}$ (i.e. $b=286$~km~s$^{-1}$). 
The expected Ly$\alpha$
absorption profile for $N_{\rm H~I}$ at various $T$
 is displayed in Figure 2,
\begin{figure}
\psfig{file=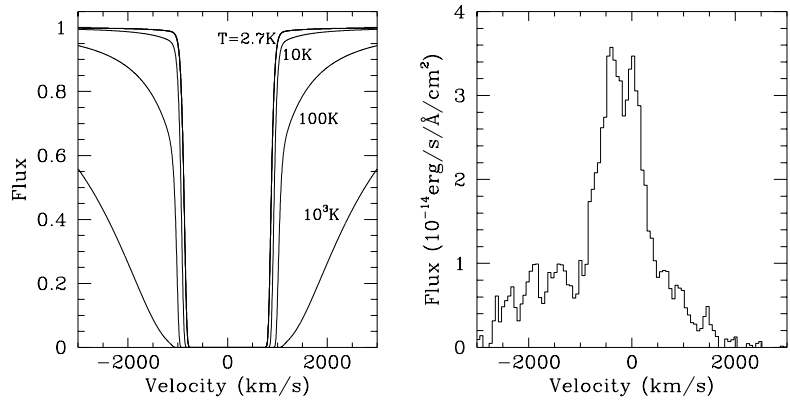,width=11.cm,angle=-90,silent=}
\caption{
The predicted vs. observed Ly$\alpha$ absorption profile.
Left: The predicted absorption profile for different values of
T with $N_{\rm H~I}=2\times 10^{18}T$ and $b=286$~km~s$^{-1}$, 
as measured by Jaffe (1990).
Right: The Ly$\alpha$ spectrum observed by Johnstone \& Fabian 
(1995). Very
little, if any, absorption is present in Ly$\alpha$.} 
\end{figure}
indicating that for all $T$ one expects a very broad absorption trough
with FWHM$\ge 2000$~km~s$^{-1}$. The Ly$\alpha$ region in NGC~1275
was observed by Johnstone \& Fabian (1995), and the observed spectrum 
is displayed in Figure 2 (velocity scale is relative to 
5260~km~s$^{-1}$). Clearly, the absorption predicted based on
the 21~cm $N_{\rm H~I}$ is not present. The small trough at the center 
of Ly$\alpha$ suggests absorption with EW$\sim 0.5$~\AA, or 
$120$~km~s$^{-1}$. Johnstone \& Fabian (1995) suggested that Ly$\alpha$ 
is double peaked, rather
than absorbed, in which case the absorption EW would be $\ll 0.5$~\AA.  

Clearly, Ly$\alpha$ implies a much lower values for
$N_{\rm H~I}$ than the 21~cm line. To obtain the $N_{\rm H~I}$ 
implied by Ly$\alpha$ one
needs to calculate EW($N_{\rm H~I}$), i.e. use the standard ``curve of
growth'' analysis. Figure 3 displays on the left hand side the
Ly$\alpha$ EW versus $N_{\rm H~I}$ for various
values of the $b$ parameter (assuming a Gaussian velocity distribution). 
The EW increases linearly with $N_{\rm H~I}$
when the line is optically thin, saturating to EW$\propto 
\sqrt{\ln N_{\rm H~I}}$ 
when the line becomes optically thick, and recovering back to 
EW$\propto \sqrt{N_{\rm H~I}}$ when the Lorenztian wings dominate the 
absorption (`damped' absorption). The observed absorption EW of 
120~km~s$^{-1}$ translates to
$10^{14}\ge N_{\rm H~I}\ge 4\times 10^{17}$~cm$^{-2}$, where the
upper limit is obtain if $b<10$~km~s$^{-1}$, and the lower limit is
obtained if $b>50$~km~s$^{-1}$. The right hand side 
curves in Figure 3 represent
the curves of growth for 21~cm absorption by H~I at $T=10$~K. These curves
are identical to those for Ly$\alpha$ absorption, but shifted by a factor
of $2.4\times 10^7$ to the right hand side. The observed 21~cm absorption
EW translates to $N_{\rm H~I}=2\times 10^{19}$~cm$^{-2}$ for a reasonable
lower limit of $T=10$~K. The largest column allowed by Ly$\alpha$
is therefore $\sim 50$ times smaller than indicated by the 21~cm line.

\begin{figure}
\psfig{file=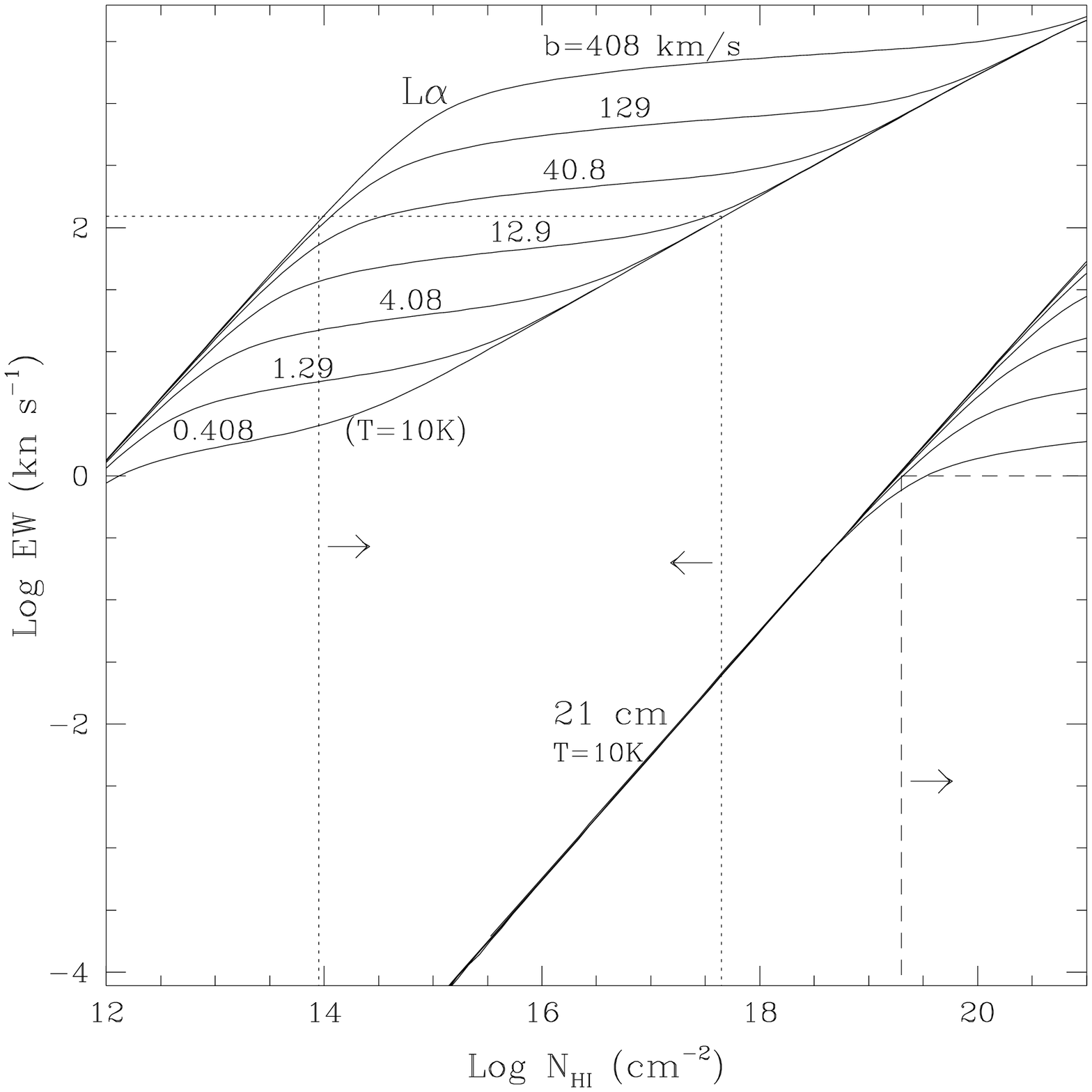,width=13.0cm,silent=}
\caption{The Ly$\alpha$ and the 21~cm curves of growth for different
velocity dispersions ($b$ parameters). The horizontal dashed lines indicate
the observed absorption EWs, and the vertical dashed lines indicate the
derived limits on $N_{\rm H~I}$. The largest column allowed by Ly$\alpha$
is $\sim 50$ times smaller than indicated by the 21~cm line.} 
\end{figure}

\section{How can the 21~cm and Ly$\alpha$ columns be reconciled?}
 
The apparent contradiction between the 21~cm and the Ly$\alpha$ columns
can be understood if the H~I column is highly non-uniform, and the 
spatial distributions of the background 21~cm and Ly$\alpha$ emission are
different. There is observational evidence that both these effects are
present in NGC~1275. According to the 21~cm continuum map presented by Jaffe 
(1990) most of the continuum originates within $\pm 20''$ of the center.
Fabian, Nulsen, \& Arnaud (1984) discovered with IUE that Ly$\alpha$ is
also extended on $\sim 10''$ scale, and the recent HUT observations by
Van Dyke Dixon, Davidsen, \& Ferguson (1996) indicate that Ly$\alpha$
emission of comparable surface brightness extends out to $\sim 60''$
from the center. Evidence for spatially non-uniform absorption is seen
on much smaller scales in the VLBA observations of Walker et al. (1994) and
Vermeulen et al. (1994), who discovered a free-free absorbed counter jet to
the north of the nucleus.
The counter jet is most likely seen through a large column disk of
relatively cold gas close to the center of NGC~1275 (Levinson et al. 1995),
while the line of sight to the southren jet is clear.
The large $N_{\rm H~I}$ indicated by the 21~cm line most likely 
resides on scales smaller than the $\sim 10-20$~kpc scale of the 
Ly$\alpha$ emitting filaments. The low
$N_{\rm H~I}$ 
indicated by Ly$\alpha$ provides a constraint for the $\sim 100-200$~kpc
scale absorber indicated by the X-ray observations.

\section{How can the X-ray and Ly$\alpha$ columns be reconciled?}

The various X-ray telescopes mentioned above indicate an excess absorbing
column of $(1.5-4)\times 10^{21}$~cm$^{-2}$, and a covering factor close
to unity. Such a column becomes optically thick
at $E<0.6-1$~keV. At this energy range O is the dominant absorber 
(e.g. Morrison 
\& McCammon 1983). Thus, the X-ray spectra merely indicate the presence
of an O column of $(1.3-3.4)\times 10^{18}$~cm$^{-2}$. The O 
X-ray absorption
is done by the inner K shell electrons, thus the ionization state of
the O can be anywhere from O~I to O~VII.

The X-ray and Ly$\alpha$ constraints imply that whatever is producing
the absorption on the $\sim 100$~kpc scale in the Perseus cooling flow 
cluster 
must have $3<N_{\rm O}/N_{\rm H~I}<3\times 10^4$, i.e. it is drastically
different from a neutral, solar abundance absorber, where
$N_{\rm O}/N_{\rm H~I}=8.5\times 10^{-4}$. 

If the absorber has roughly solar abundance then H must be highly
ionized with  $2.5\times 10^{-8}<N_{\rm H~I}/N_{\rm H}<2.5\times 10^{-4}$.
Can H be so highly ionized?  The available ionizing flux is far too
low for significant photoionization. Collisional ionization requires
$5\times 10^6>T>5\times 10^4$~K, where the upper limit prevents O from
being too
highly ionized. However, this temperature range is excluded based on
the absence of significant line emission (e.g. 
Voit \& Donahue 1995). It thus
appears that the required ionization state of H is ruled out.
The above constraints on $T$ assume equilibrium ionization states. It
remains to be studied whether plausible deviations from ionization equilibrium 
can significantly affect the ionization state of H.

Another possibility is that the absorber has practically no H. This would be
the case if the absorption originates in O which resides in dust grains embeded
in hot gas. However, the dust sputtering time scales appear too short
to explain the absorption in the inner parts of clusters (Dwek et al. 1990;
Voit \& Donahue 1995).

Could most of the H be in molecular form? 
CO emission was detected in the Perseus CF (e.g. Braine et al. 1995)
indicating that
some of the H is indeed in molecular form. However, the 
H~I/H$_2$ fraction needs to be $<2.5\times 10^{-4}$, while theoretical
calculations (Ferland et al. 1994) indicate that most 
of H ($>80$\%) would be in atomic form, even in extremely cold 
clouds embeded in CFs.

It therefore appears that there is no satisfactory model which
explains the X-ray absorption together with the new tight limits on
$N_{\rm H~I}$ obtained from  Ly$\alpha$.

\section{Future perspectives}

Given the difficulty in finding a plausible explanation for the X-ray
absorption, it is crucial to verify that this absorption is
indeed real. This can be achieved by the detection of an O bound-free 
K edge at the CF cluster redshift (see Sarazin, these proceedings). 
The edge energy will also indicate the ionization state of the absorber.
This can be achieved with next generation high resolution X-ray telescopes.

On a shorter time scale, significantly better constraints on 
$N_{\rm H~I}$ in NGC~1275 can be obtained with HST. Currently, 
the actual value of $N_{\rm H~I}$ in NGC~1275 is 
uncertain by nearly 4 orders of magnitude, and a much more accurate 
determination can be achieved through a higher resolution UV spectrum,
as demonstrated in Figure 4.
\begin{figure}
\psfig{file=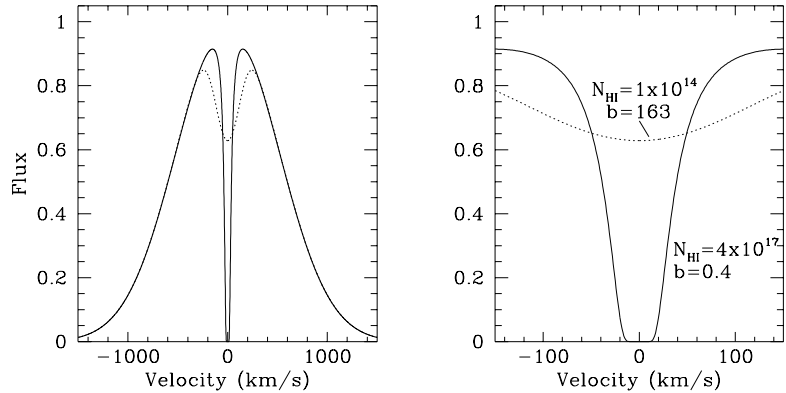,width=11.cm,angle=-90,silent=}
\caption{The expected Ly$\alpha$ profile in NGC~1275 at a high
spectral resolution ($\sim 10$~km~s$^{-1}$).  Left: 
Emission + absorption profiles. Right: Blowout of the
absorption profile. Such a spectrum would allow a rather accurate 
determination of $N_{\rm H~I}$. If the H~I gas has a very low 
velocity dispersion the absorption will go black, and if the 
velocity dispersion
is high the absorption profile will remain shallow.} 
\end{figure}
 
The method described here can be extended to many more CF clusters. All
CF clusters have a large cD galaxy at their center, and these galaxies tend
to have a power-law continuum source at their center with significant UV
emission which can be used for the detection of 
Ly$\alpha$ absorption. In addition, significant Ly$\alpha$ emission
most likely originates from the emission line filaments present in all CF
clusters. For example, Hu (1992) observed 10 CF cluster with
the IUE, detecting significant Ly$\alpha$ emission in 7 clusters, thus 
demonstrating that the method described here can be applied to most CF 
clusters.

The disadvantage of using the central cD galaxy is that if absorption
is detected it may be produced by gas local to the cD or the emission
line filaments, rather than the
large scale absorber. The lack of significant absorption can, however,
be used to place stringent limits on the nature of the X-ray absorber.

\acknowledgments

This research was partly supported by the E. and J. Bishop research fund,
and by the Milton and Lillian Edwards academic lectureship fund. The
UV spectrum of NGC~1275 was generously provided by R. M. Johnstone.

%


\begin{references}
\reference Allen, S. W., Fabian, A. C., Johnstone, R. M., Nulsen, P. E. J., 
Edge, A. C. 1992, \mnras, 254, 51
\reference Antonucci, R., \& Barvainis, R. 1994, \aj, 107, 448
\reference Braine, J., Wyrowski, F., Radford, S. J. E., Henkel, C.,
\& Lesch, H. 1995, \astap, 293, 315
\reference Dwarakanath, K. S., van Gorkom, J. H., \& Owen, F. N. 1994,
\apj, 432, 469
\reference Dwek, E., Rephaeli, Y., \& Mather, J. C. 1990, \apj, 350, 104
\reference Fabian, A. C., Arnaud, K. A., Bautz, M. W., \& Tawara, Y. 1994,
\apj, 436, L63
\reference Fabian, A. C., Nulsen, P. E. J., \& Arnaud, K. A. 1984, \mnras,
208, 179
\reference Ferland, G. J., Fabian, A. C., \& Johnstone, R. M. 1994, \mnras,
266, 399
\reference Jaffe, W. 1990, \astap, 240, 254
\reference Johnstone, R. M., \& Fabian, A. C. 1995, \mnras, 273, 625
\reference Hu, E. M. 1992, \apj, 391, 608
\reference Levinson, A., Laor, A., \& Vermeulen, R. C. 1995, \apj, 448, 589
\reference Morrison, R., \& McCammon, D. 1983, \apj, 270, 119
\reference O'Dea, P. C., Baum, S. A., Maloney, P. R., Tacconi, L. J., 
\& Sparks, W. 1994, \apj, 422, 467
\reference Van Dyke Dixon, W., Davidsen, A. F., \& Ferguson, H. C. 1996,
\aj, 111, 130
\reference White, D. A., Fabian, A. C., Johnstone, R. M., Mushotzky, R. F.,
\& Arnaud, K. A. 1991, \mnras, 252, 72
\reference Vermeulen, R. C., Readhead, A. C. S., \& Backer, D. C. \apj,
430, L41
\reference Voit, G. M., \& Donahue, M. 1995, \apj, 452, 164
\reference Walker, R. C., Romney, J. D., \& Benson, J. M. 1994, \apj 430, L45


\end{references}
\end{document}